\documentclass[prb,amsfonts,amssymb,twocolumn,floats,superscriptaddress,aps]{revtex4-2}
\usepackage{setspace}
\usepackage[latin1]{inputenc}
\usepackage{graphicx}
\usepackage{physics}
\usepackage{amssymb,amsmath,braket}
\usepackage{upgreek}
\usepackage{bm,bbold}
\usepackage[dvipsnames]{xcolor}
\usepackage{graphicx,tikz}
\usepackage[normalem]{ulem}
\usepackage{color, colortbl}
\usepackage{mathbbol}
\usepackage{bbm}
\usepackage{tikz}

\usepackage{hhline}
\usepackage{array}
\usepackage{float}
\usepackage{subfigure}
\usepackage{multirow}

\usepackage[colorlinks,linktocpage]{hyperref}
\hypersetup{
	citecolor  = NavyBlue,
	linkcolor  = NavyBlue,
	urlcolor = NavyBlue
}

\DeclareMathOperator{\sgn}{sgn}
\DeclareMathOperator{\re}{Re}
\DeclareMathOperator{\im}{Im}

\newcommand{\iu}{\mathrm{i}}
\newcommand{\eu}{\mathrm{e}}

\newcommand{\be}{\begin{equation}}
\newcommand{\ee}{\end{equation}}
\newcommand{\beq}{\begin{equation}}
\newcommand{\eeq}{\end{equation}}
\newcommand{\bea} {\begin{eqnarray}}
\newcommand{\eea} {\end{eqnarray}}

\newcommand*\diff{\mathop{}\!\mathrm{d}}

\newcommand{\ve}[1]{{\bf #1}}

\newcommand{\kv}{\ve{k}}
\newcommand{\qv}{\ve{q}}
\newcommand{\Qv}{\ve{Q}}

\newcommand{\Dpg}{\Delta_{\rm pg}}
\newcommand{\tDpg}{\tilde{\Delta}_{\rm pg}}

\newcommand{\bw}{\bar{\omega}}

\definecolor{bananayellow}{rgb}{1.0, 0.88, 0.21}
\definecolor{straw}{rgb}{0.32, 0.28, 0.1}

\begin{document}
\title{Crucial role of thermal fluctuations and vertex corrections for the magnetic pseudogap}
\author{Mengxing Ye}
\affiliation{Kavli Institute for Theoretical Physics, University of California, Santa Barbara, CA 93106, USA}
\affiliation{Department of Physics and Astronomy, University of Utah, Salt Lake City, UT 84112, USA}
\author{Andrey V Chubukov}
\affiliation{School of Physics and Astronomy and William I. Fine Theoretical Physics Institute, University of
Minnesota, Minneapolis, MN 55455, USA}
\begin{abstract}
It is generally believed that in a  2D metal, whose ground state is antiferromagnetically ordered with ${\bf Q} = (\pi,\pi)$, thermal (static) magnetic fluctuations give rise to precursor behavior above $T_N$, in which the spectral function of a hot fermion (the one for which ${\bf k}$ and ${\bf k} + {\bf Q}$ are Fermi surface points) contains two peaks, separated by roughly the same energy as in the antiferromagnetically ordered state. The two peaks persist in some range of  $T >T_N$ and eventually merge into a single peak at zero frequency. This behavior is obtained theoretically by departing from free fermions in a paramagnet and evaluating the dressed fermionic Green's function by summing up infinite series of diagrams with contributions from thermal magnetic fluctuations.
We show, following [Y.M. Vilk and A.-M. S. Tremblay, J. Phys. I France ${\bf 7}$ 1309 (1997)] that keeping vertex renormalization diagrams
in these series is crucial as other terms only broaden the spectral function of a hot fermion, but do not shift its maximum away from zero frequency.  As the consequence, the magnetic
pseudogap should be treated as an input for theories that neglect vertex corrections, like, e.g., Eliashberg theory for magnetically-mediated superconductivity.
We also analyze the potential pseudogap behavior at $T=0$. We argue that it may exist, but only at a finite correlation length, and not as a precursor to antiferromagnetism.
\end{abstract}
\date{\today}
\maketitle

{\it \bf Introduction.}~~~The origin of the pseudogap behavior, observed in the cuprates and other correlated materials
 remains  the  subject of
ongoing debate.  Theoretical proposals for the pseudogap can be broadly split into three categories. One identifies
 pseudogap behavior with some particle-hole order, either a conventional one,
like a charge-density wave (CDW)~\cite{Metlitski2010,WangYuxuan2014,Chowdhury2014,Atkinson2015,Grilli}, or less conventional, like a circulating current~\cite{Varma1997,Varma1999}. Another identifies the pseudogap with a spin-liquid-type ``mother" state, from which one gets antiferromagnetism, superconductivity, and charge order~\cite{Sachdev2019,YHZhang2020a,Sachdev2022pga,Sachdev2022pgb,Sachdev202302}. And the third treats the pseudogap phase as a precursor to an ordered state - either  a spin-density-wave (SDW)
order
~\cite{Vilk1996,Vilk1997,Schmalian1998,Schmalian1999,Sadovskii1999,Moca2000,Sadovskii_review,Yanase2004,Roy_2008,Sedrakyan2010,Gull2015,Gunnarsson2015,
Ye2019pg,Schafer2021,Held2022,Krien2021,
Simkovic2022,*Simkovic2022a,Ye2023},  or superconductivity~\cite{Randeria1998,Millis1998,Fujimoto2002,Yanase2004,Berg2007,YMWu2021,Dai2021,Qi2022},
 or pair-density-wave~\cite{Dai2020}.

In this communication we focus on the last category and discuss some aspects of a precursor to an SDW order with ${\bf Q} = (\pi,\pi)$  in two dimensions.  We analyze the emergence  of  peaks at a finite frequency in the spectral function of a fermion on the Fermi surface, particularly at a hot spot ${\bf k}_{hs}$, for which ${\bf k}_{hs}$ and ${\bf k}_{hs} + {\bf Q}$ are both
  on the Fermi surface. The emergence of such peaks without a full gap between them is a signature
   feature of pseudogap behavior.

We address two issues.  The first is
 about pseudogap behavior
 caused by thermal magnetic fluctuations~
\cite{Vilk1997,Schmalian1999,Sadovskii1999,Moca2000,Sadovskii_review,Yanase2004,Sedrakyan2010,Gull2015,Ye2019pg,Schafer2021,Held2022,Krien2021,Ye2023}.
Several groups, including us, demonstrated~\cite{Schmalian1999,Sadovskii1999,Sadovskii_review,Yanase2004,Sedrakyan2010,Ye2019pg,Ye2023} that
 that pseudogap does develop when one includes infinite series of contributions to the fermionic Green's function  from   thermal (static) spin fluctuations.
In this communication, we look more closely
 at the interplay between non-crossed and crossed diagrams in these series.
  The non-crossed diagrams  renormalize the Green's function of an intermediate fermion,
   $G_0 (\kv+\qv, \omega_m) \to G(\kv+\qv, \omega_m)$, and can be absorbed into the self-consistent one-loop theory (SCOLT).  The crossed diagrams describe vertex corrections.
  Previous studies~\cite{Vilk1996,Vilk1997,Moca2000,Schafer2021} found that at large dimensionless spin-fermion
   coupling $\lambda_{\rm th}$, the non-crossed diagrams, taken alone, broaden the spectral function of a hot fermion, but
   the maximum  remains at $\omega =0$, i.e., pseudogap does not emerge.
Here, we show that (i)  pseudogap behavior does not develop within SCOLT for any value of $\lambda_{\rm th}$, (ii) SCOLT  is the ``boundary" case in the sense that     already infinitesimally small vertex corrections
      give rise to a pseudogap, and (iii)       SCOLT is a member of a one-parameter set of such boundary models,
 which do not display  pseudogap behavior, but  develop it
 upon an infinitesimally small perturbation.

Second, we analyze whether  the system can potentially display pseudogap behavior  at $T=0$.
  We argue that this holds
 in the weak coupling regime away from the SDW quantum-critical point (QCP), when SDW fluctuations are gapped and weakly damped.
 Close to the SDW QCP,  Landau damping takes over and $A({\bf k}_{hs}, \omega)$ has a maximum at $\omega =0$.  This agrees with
  the recent study by Grossman and Berg~\cite{Berg2023}. In a generic case when
  fermionic velocity $v_F$ and bosonic velocity $v_s$ are comparable,
  pseudogap behavior ends up
   when the system enters the strong coupling regime near a QCP.
      If, however, $v_s$ is small compared to  $v_F$, pseudogap behavior
       extends into the strong coupling regime.
       We emphasize that this pseudogap is not a precursor to SDW as the magnitude of the pseudogap in $A({\bf k}_{hs}, \omega)$ is set by the mass of the SDW fluctuations, and it must disappear before a QCP.

{\it \bf Pseudogap due to thermal fluctuations}~~~We consider a system of fermions, interacting by exchanging spin fluctuations with momentum near ${\bf Q}$. We take as an input that static spin fluctuations have Ornstein-Zernike form  with a large, but finite correlation length $\xi = \xi (T)$ and are coupled to fermions by  Yukawa coupling ${\bar g}$, which we assume to be comparable to the bandwidth.   Our goal is to obtain the spectral function
  $A(\kv_{\rm hs},\omega)=-(1/\pi)$ Im$G_{\text{ret}} (\kv_{\rm hs},\omega)$
 for a hot fermion and verify whether at a finite $T$ and large, but still finite $\xi$, its maximum splits into two maxima at a finite frequency, and whether vertex corrections are crucial for the spitting.
 For this specific goal, it is sufficient to treat $\xi = \xi (T)$ as an input parameter (for self-consistent calculations of $\xi (T)$ see Refs.~\cite{Roy_2008,Ye2023}).

The spectral function generally can easily obtained by evaluating the thermal self-energy $\Sigma_{\rm th}({\bf k}_{\rm hs},\omega)$.
We first compute it at the one-loop order, use the result to rationalize the need to include higher-loop contributions, and then analyze
 $A(\kv_{\rm hs},\omega)$
 and the dressed $\Sigma_{\rm th}({\bf k}_{\rm hs},\omega)$ with and without vertex corrections.

\begin{figure}
\includegraphics[width=0.85\columnwidth]{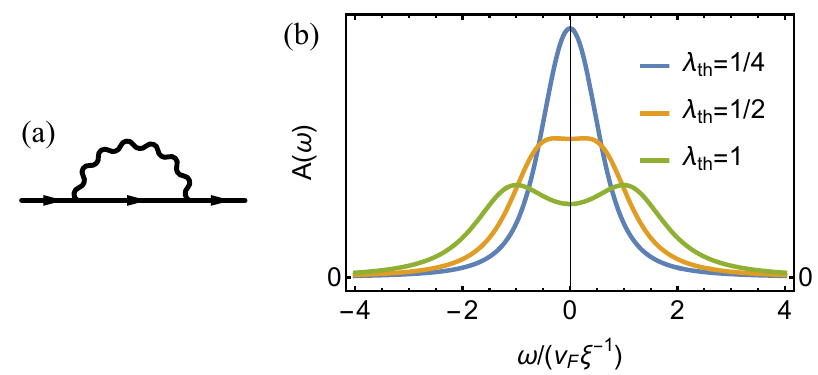}
\caption{(a) One-loop self-energy. (b) Spectral function at the hot spot from the one-loop calculation. As the dimensionless
coupling $\lambda_{\rm th} = \frac{3\bar{g}T}{2\pi (v_F \xi^{-1})^2} $ increases, the spectral function shows pseudogap
behavior when $\lambda_{\rm th}>\lambda_c=0.47$.
\label{fig:oneloop}}
\end{figure}

The one-loop thermal
 self-energy, shown in Fig.~\ref{fig:oneloop}~(a), is the convolution of a propagator  of a free  fermion,  $G^{(0)} ({\bf k}_{\rm hs} + {\bf Q} +{\bf q}, \omega)$,  and a static spin propagator $\chi (\qv) = 1/(\qv^2 + \xi^{-2})$. Expanding the fermionic dispersion to linear order in ${\bf q}$ and integrating over the two components of ${\bf q}$,
 one obtains the exact analytical expression~\cite{Vilk1997,Moca2000,Roy_2008,Schafer2021,Ye2023}
\bea
&&\Sigma^{(1)}_{\rm th}({\bf k}_{\rm hs}, \omega)=  v_F \xi^{-1} \lambda_{\rm th} \times \nonumber \\
&& \left[\text{sign} (\mathsf{w}) \frac{\log \left(\mathsf{w}+\sqrt{(\mathsf{w})^2+1} \right) }{{\sqrt{(\mathsf{w})^2+1}}} - \iu \frac{\pi}{2 \sqrt{(\mathsf{w})^2+1}}\right]
\label{eq:oneloop0}
\eea
 where $\lambda_{\rm th}
 = (3\bar{g}T(2\pi (v_F \xi^{-1})^2)$ is the
dimensionless ``thermal"  coupling, and
$\mathsf{w}=\omega/(v_F \xi^{-1})$ is the dimensionless frequency.
The dimensionless coupling grows as the system approaches the onset temperature $T_N$ of the $(\pi,\pi)$ order.

We show the spectral function $A^{(1)}({\bf k}_{hs}, \mathsf{w}) = -(1/\pi)\im \left[\left(v_F \xi^{-1}  \mathsf{w} - \Sigma^{(1)}_{\rm th}({\bf k}_{\rm hs}, \mathsf{w})\right)^{-1}\right]$  in Fig.~\ref{fig:oneloop} (b).
At small $\lambda_{\rm th}$, $A^{(1)}(\kv_{\rm hs},\mathsf{w})$ is peaked at $\mathsf{w} =0$, as is expected for a weakly
interacting fermion at the Fermi surface.
However once $\lambda_{\rm th}$ exceeds the critical value $\lambda_c = 2\sqrt{2}/(2\sqrt{2} + \pi) \approx 0.4738$, the maximum of $A^{(1)} (\kv_{\rm hs},\mathsf{w})$
 shifts to a finite $|\mathsf{w}|  = \tDpg \sim \sqrt{\lambda - \lambda_c}$.
The pseudogap behavior becomes particularly
 pronounced at large $\lambda_{\rm th}$, where
 $\tDpg > v_F \xi^{-1}$, and
 at $\omega \sim \tDpg$,
\beq
\int_{\qv}\, G^{(0)} (\kv_{\rm hs} + {\bf Q} + {\bf q}, \omega)  \chi (\qv) \approx  G^{(0)} (\kv_{\rm hs} + {\bf Q}, \omega) \int_{\qv}\, \chi (\qv)
\label{ll_1}
 \eeq
 such that
  $\Sigma^{(1)}_{\rm th} (\kv_{\rm hs}, \omega) \approx \tDpg^2/\omega$~\cite{Vilk1997,Moca2000,Roy_2008,Schafer2021,Ye2023}
  with
  $\tDpg = (v_F \xi^{-1}) (\lambda \log{\lambda})^{1/2}/\sqrt{2} \approx \left(\frac{3{\bar g} T}
 {2\pi}  \log{\xi}\right)^{1/2}$. Here, $\int_\qv=\int \diff^2 \qv $.
 This self-energy is the same as  in the SDW-ordered state, hence the emergence of the peaks at $|\omega| = \tDpg$ is quite natural
(below the peak, $\im \Sigma^{(1)}(\kv_{\rm hs}, \omega)$ remains non-zero down to the lowest frequencies, hence $\tDpg$ is a pseudogap rather than a true gap).

\begin{figure}
\includegraphics[width=0.8\columnwidth]{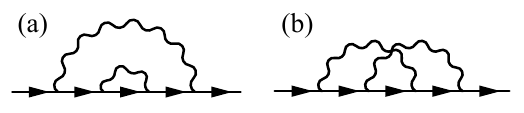}
\caption{
Non-crossed (a) and crossed (b) two-loop irreducible diagrams.
\label{fig:twoloop}}
\end{figure}

 We see that  the pseudogap behavior
  does emerge within the one-loop approximation, however the coupling $\lambda_{\rm th}$ must exceed $\lambda_c = O(1)$.
  This  raises the question whether the one-loop result  stands once we include higher-order terms.
   Examples of higher-order diagrams for $\Sigma (\kv_{\rm hs}, \omega)$ are shown in Fig.~\ref{fig:twoloop}.
  They include non-crossed diagrams (Fig.~\ref{fig:twoloop}~(a)), which account for the renormalization of the internal fermionic line, and crossed diagrams (Fig.~\ref{fig:twoloop}~(b)), which represent  vertex corrections.
 Besides, the chemical potential $\mu$ is different from $\mu_0 = \epsilon_{{\bf k}_{hs}}$ and is obtained self-consistently from the condition on the fermionic density.
 Below we incorporate the change of chemical potential into
$\omega \to {\bar \omega} = \omega + \delta \mu$, where $\delta \mu =\mu-\epsilon_{\kv_{hs}}=\mu-\mu_0$.

As a first step, let's
 keep only non-crossed higher-loop diagrams, i.e., neglect vertex corrections.
  The fully dressed self-energy is  given by the
   same one-loop diagram as in the perturbation theory, but with the fully dressed propagator of an intermediate fermion. This is the SCOLT.
   The retarded Green's function is $G^{sc} (\kv_{\rm hs}, \mathsf{\bar w})^{-1} =  v_F \xi^{-1}X$ (sc stand for self-consistent), where
   $\mathsf{\bar w} = {\bar \omega}/v_F \xi^{-1}$, and  $X = X(\mathsf{\bar w})$ is the solution of
  \beq
  X = \mathsf{\bar{w}} - \lambda_{\rm th} \frac{\log\left(X + \sqrt{X^2 +1}\right)}{\sqrt{X^2+1}}
 + \iu \lambda_{\rm th} \frac{\pi}{2\sqrt{X^2+1}}
\label{ff_1}
\eeq
 Expanding at small $\mathsf{\bar w}$, we find
  [see Supplementary material (SM) for detail],
  $X = a\mathsf{\bar w} + ib (1 - c \mathsf{\bar w}^2) +...$, where $a,b$ and $c$ are functions of $\lambda_{\rm th}$ and dots stand for terms of higher order in $\mathsf{\bar w}$.  The spectral function $A(\kv_{\rm hs},\mathsf{\bar w}) \propto 1/(b^2 + \mathsf{\bar w}^2 (a^2-b^2 c))$.
 The pseudogap emerges when the prefactor for $\mathsf{\bar w}^2$ is negative, i.e., when $a^2 < b^2 c$.  We expanded analytically in $\mathsf{\bar w}^2$ and found
  that this does not happen at any value of  $\lambda_{\rm th}$:
    the quasiparticle peak broadens as $\lambda_{\rm th}$ increases, but remains centered at $\omega =0$.  At large $\lambda_{\rm th}$,
  when $\tDpg > v_F \xi^{-1}$, the spectral function
  has a semi-circular form
  $A^{sc}(\kv_{\rm hs}, {\bar \omega}) = \sqrt{4 \tDpg^2 - {\bar \omega}^2}/(2\pi \tDpg^2)$ at
   ${2\tDpg > \bar \omega} > v_F \xi^{-1}$~\cite{Vilk1997} and remains smooth at $\omega < v_F \xi^{-1}$ as we verified.
   This spectral function describes incoherent excitations extending up to $2 \tDpg$,
   and its maximum remains at $\omega =0$
    \footnote{We note in passing that the total spectral weight is the same as in $A^{(1)}(\kv_{\rm hs}, {\bar \omega})$, where at such $\lambda_{\rm th}$ it is concentrated around the near-$\delta$- functional peaks at $|{\bar \omega}| =\tDpg$.}.

\begin{figure}
\includegraphics[width=1\columnwidth]{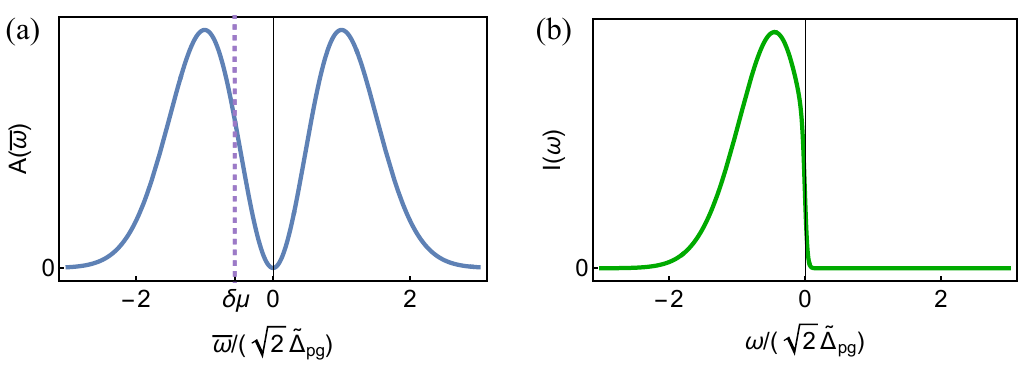}
\caption{(a) Spectral function $A^{\rm full}(\kv_{hs}, \bar{\omega})$ and (b) Spectral intensity  $I^{\rm full}(\kv_{hs}, \omega)$ for the SU(2) symmetric model (see Eqs.~\eqref{ll}).
 The horizontal axis is
 $\bar{\omega}=\omega+\delta \mu$ in (a) and
  $\omega$ in (b),  both in units of $\sqrt{2}\tDpg$.
\label{fig:Spectral}}
\end{figure}

We next include the crossed diagrams.
 We compute the full self-energy directly, by extending perturbation theory to infinite order~\cite{Sadovskii1974,*Sadovskii1974b,*Sadovskii1979,*SadovskiiBook}.
The computations again simplify at large $\lambda_{\rm th}$, where we can use Eq.~\eqref{ll_1}. Using it for all diagrams, we
 find that at each loop order the crossed and the non-crossed diagrams are of the same order, and each  set
 forms series in $G^{(0)} (\kv_{\rm hs} + {\bf Q}, {\bar \omega}) G^{(0)} (\kv_{\rm hs}, {\bar \omega}) \tDpg^2$. This allows one to keep only one diagram at a given loop order $m$ and multiply it by
  the proper combinatoric  factor $\mathcal{D}_m$.    For the SU(2)-symmetric problem, $\mathcal{D}_m = (2m+1)!!$~\cite{Schmalian1999,Sedrakyan2010}.
  The full Green's function is then
  $G^{full}(\kv_{\rm hs},{\bar \omega}) = G^{(0)}(\kv_{\rm hs}, {\bar \omega}) C(\bar \omega)$, where
\beq
C(\bar \omega) = \sum_m
(2m+1)!! \left(\tDpg^2 G^{(0)}(\kv_{\rm hs}, {\bar \omega})  G^{(0)}(\kv_{\rm hs}+\Qv, {\bar \omega}) \right)^m
\label{eq:twopoint2}
\eeq
 Re-expressing the infinite sum
 as the integral
\begin{equation}
C ({\bar \omega}) = \frac{2}{\sqrt{\pi}}\int_0^\infty \diff t \, \eu^{-t} \frac{t^{1/2}}{1- u t}
\label{eq:fullGreen}
\end{equation}
where
$u = 2\tDpg^2 G^{(0)}(\kv_{\rm hs}, {\bar \omega}) G^{(0)}(\kv_{\rm hs}+\Qv,{\bar \omega})$, and
 evaluating it,
we obtain for a fermion at a hot spot
\beq
 C(\bw) = \mathsf{C}(z) = 2 z^2 \left(\left(\sqrt{\pi} z \eu^{-z^2} {\rm{Erfi}}(z)-1\right) - i \sqrt{\pi} z \eu^{-z^2}\right)
 \eeq
where $z = {\bar \omega}/(\tDpg \sqrt{2})$, and  ${\rm Erfi}(z)$ is the imaginary error function.
 The spectral function is
\be
A^{full}(\kv_{\rm hs},{\bar \omega})=\frac{1}{\sqrt{2 \pi}\tDpg}\frac{{\bar \omega}^2}{\tDpg^2}\exp\left[-\frac{\bw^2}{2\tDpg^2}\right]
\label{ll}
\ee
We plot the full spectral function in Fig.~\ref{fig:Spectral}~(a). We see that it does display the pseudogap behavior.  The form of the full $A^{full}(\kv_{\rm hs},\bw)$  is rather similar to the one-loop result at $\lambda_{\rm th} \gg 1$, and the
value of the full pseudogap $\Dpg$
 is comparable to $\tDpg$
\footnote{We note in passing that within SCOLT, the Green's function can also be represented as an infinite sum of multi-loop diagrams, like in Eq. (\ref{eq:twopoint2}), but with the combinatoric factor  $\mathcal{D}_m = 2^{2m+1} (2m-1)!!/(2m+2)!!$. Solving Eq.~\eqref{eq:twopoint2} with this $\mathcal{D}_m$, we reproduce the  spectral function solution $A^{(sc)}(\kv_{\rm hs},\bw) \propto \sqrt{4\tDpg^2-\bw^2}$.}.

\begin{figure*}
\subfigure[]{\includegraphics[width=0.48\columnwidth]{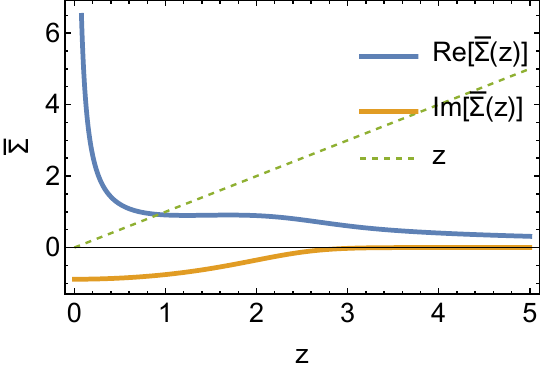}}
\subfigure[]{\includegraphics[width=0.48\columnwidth]{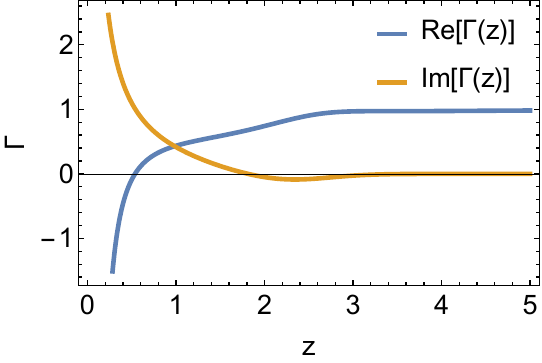}}
\subfigure[]{\includegraphics[width=0.48\columnwidth]{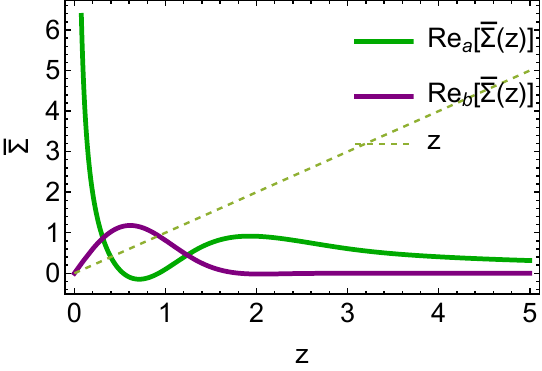}}
\subfigure[]{\includegraphics[width=0.5\columnwidth]{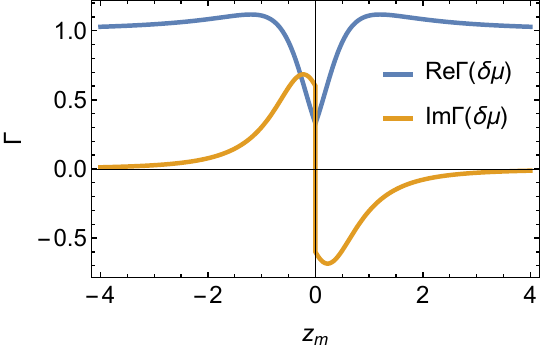}}
\caption{Panels (a) and (b): real and imaginary part of the normalized self-energy (a) and  the vertex function (b) as functions of $z = (\omega + \delta \mu)/(\tDpg \sqrt{2})$, from Eq.~\eqref{eq:nSG}. The
the spectral function has a peak at $z \approx 1$, where $\re {\bar \Sigma} (z)$ crosses $z$.
Panel (c): two components of $\re{\bar \Sigma} (z)$: $ \re_a = (3/2z) = \mathsf{C} (z)\re\Gamma (z)$ and $\re_b=-(3/2z) \im\mathsf{C} (z)\im\Gamma (z)$. Near $z=1$, $\re{\bar \Sigma} (z) \approx z \approx -(3/2z) \im\mathsf{C} (z)\im\Gamma (z)$.
Panel (d): $\re \Gamma (z_m)$ and $\im \Gamma (z_m)$ along the Matsubara axis at $\delta \mu =-0.8$}
\label{fig:Vertex}
\end{figure*}

A complimentary way to understand the  importance of vertex corrections  is to analyze the structure of the thermal self-energy. Dyson equation expresses it in terms of the full Green's function $G^{full} ({\bf k}_{hs}+\Qv, \bw)$ and the full vertex $\Gamma ({\bf k}_{hs}, \bw)$ as
\beq
\Sigma_{\rm th} ({\bf k}_{hs}, \bw) =
3 \tDpg^2 G^{full} ({\bf k}_{hs}+\Qv, \bw) \Gamma ({\bf k}_{hs}, \bw).
\label{ex_1}
\eeq
In the SCOLT, $\Gamma ({\bf k}_{hs}, \bw) =1$.
Using $G^{full} =  (G^{(0)} - \Sigma)^{-1}$, we find
\bea
&& \Sigma_{\rm th} ({\bf k}_{hs}, \bw) =  \sqrt{2}\tDpg  {\bar \Sigma} (z),~ {\bar \Sigma} (z)=  \frac{z (\mathsf{C}(z)-1)}{\mathsf{C}(z)}   \nonumber \\
&& \Gamma ({\bf k}_{hs}, \bw) = \Gamma (z) = \frac{2}{3} z^2 \frac{\mathsf{C}(z)-1}{\mathsf{C}^2(z)}
\label{eq:nSG}
\eea
Because $\mathsf{C}(z)$ is complex, ${\bar \Sigma} (z)$ and $\Gamma (z)$ are complex functions of the frequency.
 The spectral function is $A^{full}(\kv_{hs},\bw) = A^{full} (z)$, where
\beq
 A^{full} (z) = \left(-\frac{1}{\pi\sqrt{2} \tDpg}\right) \frac{\Im \bar{\Sigma} (z)}{(z - \Re\bar \Sigma (z))^2 + (\Im \bar \Sigma (z))^2}
\eeq
We plot real and imaginary parts of ${\bar \Sigma} (z)$ and $\Gamma (z)$ in Fig.~\ref{fig:Vertex} (a,b).
  We see that Im${\bar \Sigma} (z)$  is a rather smooth function of $z$ and is featureless around $z=1$, where the spectral function has a pseudogap peak (see Fig.~\ref{fig:Spectral}~(a)).
   On more careful look, we find that the peak in $A^{full} (z)$ at $z =1$ emerges
    because
     $\re{\bar \Sigma} (z)-z$ changes sign very near $z=1$ (see Fig.~\ref{fig:Vertex} (a)).
   Further,
  $ \re {\bar \Sigma} (z) = (3/(2z)   \left(\re \mathsf{C} (z)\re\Gamma (z)-\im\mathsf{C} (z)\im\Gamma (z)\right)$.
  We plot the two parts of this expression separately in Fig.~\ref{fig:Vertex} (c).
  We see that near $z=1$,
  $\im\mathsf{C} (z)\im\Gamma (z) \gg \re \mathsf{C} (z)\re\Gamma (z)$.
  This implies that the imaginary part of the vertex $\Gamma (z)$
  is crucial for the pseudogap. One could not obtain the peak in $A^{full} (z)$ if $\Gamma (z)$ was a constant,
  like
  in the SCOLT.

We note in passing that this analysis is
different from the one in Refs.~\cite{Held2022,Krien2021}. These authors analyzed the vertex function on the Matsubara axis.  The latter is complex at a hot spot  due to a finite $\delta \mu$, which makes even $G^{(0)} ({\bf k}_{hs}. \omega_m) = 1/(i\omega_m + \delta \mu)$ complex (Refs. \cite{Gu2020,Georges2001,WangYX2020}).  In Fig.~\ref{fig:Vertex} (d), we plot the real and imaginary parts of $\Gamma (\omega_m)$ for $\delta \mu =0$ (dashed lines) and $\delta \mu =-0.8$ (solid lines) in unit of $\sqrt{2} \tDpg$.  We see that $\im \Gamma(z_m)$
    is finite for $\delta \mu \neq 0$.
The behavior of $\re \Gamma(z_m), \im \Gamma(z_m)$ for $\delta \mu =-0.8$ is quite similar to the vertex function extracted from the numerical analysis of the self-energy in Ref. \cite{Held2022,Krien2021}.   At the same time, our results do not support the key point of \cite{Held2022,Krien2021} that the complex structure of $\Gamma (\omega_m)$ on the Matsubara axis is the key to pseudogap development. Indeed,
on the real axis, $\delta \mu$ shifts the frequency $\omega $ to ${\bar \omega}$, but the two-peak pseudogap behavior
emerges independent on the value of $\delta \mu$ and would hold even if $\delta \mu$ was zero
\footnote{A finite negative $\delta \mu$ relates the observable photoemission intensity $I({\bf k}_{hs}, \omega) = A^{full} ({\bf k}_{hs}, \omega) n_F (\omega)$ to the behavior of $A^{full} ({\bf k}_{hs}, {\bar \omega})$ at negative ${\bar \omega}$, subject to  ${\bar \omega}<- |\delta \mu|$.
Because of this constraint, the singular behavior of $\Sigma_{\rm th} (z)$ and $\Gamma (z)$ at $z \to 0$ (Fig. \protect\ref{fig:Vertex}d) is not  accessible in photoemission experiments.}.
A similar behavior of  vertex function $\Gamma$ in real and imaginary frequency has been observed in Ref.~\cite{vanLoon2018} using dynamical mean field theory.

\begin{figure}
\includegraphics[width=0.75\columnwidth]{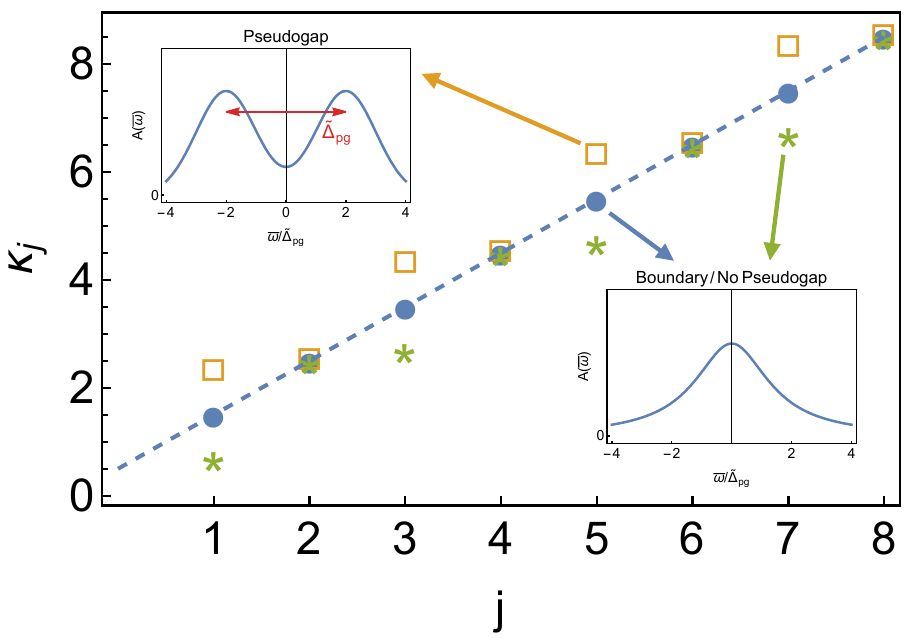}
\caption{Ilustration that the set of models with $\kappa_j = \kappa^{(0)} + \kappa^{(1)} j$  (see Eq.~\eqref{eq:condfraction})
 are at the boundary of the pseudogap formaton. The boundary models from the set are along the solid line. We verified that the pseudogap emerges at infinitesimally small $\delta k >0$ at odd $j$ (orange square).
\label{fig:kappaj}}
\end{figure}

On a more careful look, we found
that not all diagrammatic series with both non-crossed and crossed diagrams
lead to pseudogap behavior.
An example is series with the
combinatoric factor $\mathcal{D}_m=(2m-1)!!$, which holds in certain 1D models~\cite{Sadovskii1974,*Sadovskii1974b,*Sadovskii1979,*SadovskiiBook}
and 2D models  on a triangular lattice~\cite{Ye2019pg}.
These series yield $A^{\rm full}(\kv_{\rm hs},\omega) \propto \exp\left( -\bw^2/(2 \tDpg^2)\right)$, which is peaked at $\omega =0$.
 For a generic  $\mathcal{D}_m$, the series can be represented as a continued fraction
\begin{equation}
G^{full}(\kv_{\rm hs},\iu \omega_n)=\frac{1}{\iu \omega_n - \frac{\kappa_1 \tDpg^2}{\iu \omega_n - \frac{\kappa_2 \tDpg^2}{\iu \omega_n - \frac{\kappa_3 \tDpg^2}{\iu \omega_n - ...}}}},
\label{eq:condfraction}
\end{equation}
We find  that for a set of models with $\kappa_j =
\kappa^{(0)} + \kappa^{(1)} j$, the spectral function does not show pseudogap behavior. The SCOLT is a member of this set with $  \kappa^{(0)} =1$ and $ \kappa^{(1)} =0$.
 The case $ \kappa^{(0)}=0,  \kappa^{(1)}=1$ corresponds to $\mathcal{D}_m=(2m-1)!!$.
We verified numerically that for
 each member of this set,
an infinitesimally small deviation $\delta \kappa >0 $ for odd $j$
 leads to pseudogap formation (see Fig.~\ref{fig:kappaj}).
   For the model with  $
   \kappa_j =  \kappa^{(1)} j$ we
    found analytically
\begin{equation}
A^{\kappa}(\kv_{\rm hs}, \omega) \propto \abs{\omega}^{\delta \kappa /  \kappa^{(1)}} \eu^{-\frac{\omega^2}{2 \kappa^{(1)} \tDpg^2}}
\end{equation}
This spectral function has two peaks at $|\omega| = \sqrt{\delta \kappa} \tDpg$.

That SCOLT is the boundary case for the pseudogap formation
can also be seen by analyzing  a simple toy model~\cite{Sadovskii1999GL,Sadovskii2000}, in which the self-energy at large $\lambda$ is given by
\beq
\Sigma^{\rm toy} (\kv_{\rm hs}, \omega) = \tDpg^2 \left(\alpha G(\kv_{\rm hs}, \omega)  + (1-\alpha) G^{(0)}(\kv_{\rm hs}, \omega)\right)
\label{eq:Stoy}
\eeq
where $0 \leq \alpha \leq 1$.
This self-energy interpolates between perturbative one-loop theory at $\alpha =0$ and  SCOLT at $\alpha =1$.  The spectral function $A^{\rm toy} (\kv_{\rm hs}, \omega)$ is readily obtained by solving  the self-consistent equation for the Green's function
 $ G^{-1}(\kv_{\rm hs}, \omega) = \omega - \tDpg^2 \left(\alpha G(\kv_{\rm hs}, \omega)  + (1-\alpha) G^{(0)}(\kv_{\rm hs}, \omega)\right)$
 [see SM for detail].   For any $\alpha <1$, the maximum of $A^{\rm toy} (\kv_{\rm hs}, \omega)$ is at a finite
 $|\omega|= \tDpg (1-\alpha)^{1/2}$,  at $\alpha =1$ it is at $\omega =0$.
We again see that the SCOLT is the boundary case for the pseudogap formation.

{\it \bf Pseudogap from quantum fluctuations.}~~~
 We argued above that thermal spin fluctuations give rise to pseudogap behavior as a precursor to the $(\pi,\pi)$ ordered state.  We now contrast this behavior with the one  at $T=0$. We neglect superconductivity and analyze whether quantum spin fluctuations can give rise to the pseudogap.

  We use the same model as before, but with the dynamical spin propagator
    $\chi (\ve{q},\Omega_m)=\chi_0/((\Omega_m/v_s)^2+(\ve{q}-\ve{Q})^2+\xi^{-2} + \gamma |\Omega_m|)$, where $v_s$ is spin velocity and
     the last term is the Landau damping with
     $\gamma = (4/\pi \sin{\theta}) {\bar g}/v^2_F$, where $\theta$ is the angle between Fermi velocities at $\kv_{hs}$ and $\kv_{hs} + {\bf Q}$ ~\cite{Abanov2003}.
      We restrict with perturbative one-loop analysis as higher-loop terms at $T=0$ are at most $O(1)$
       relative to the one-loop term~\cite{Abanov2003}.
The exact one-loop self-energy can be readily obtained (see SM for detail),
 and its analysis shows that at  small $\lambda_{\rm q}=3 {\bar g}/(4\pi v_F \xi^{-1})$,
   the spectral function $A^{\rm q}
   ({\bf k}_{hs}, \omega)$
    nearly vanishes at  $|\omega| < v_s \xi^{-1}$ and
    has a peak at  $|\omega| \geq v_s \xi^{-1}$.
   In the opposite limit of large $\lambda_{\rm q}$, the Landau damping term is the strongest one in the spin propagator,
    and $A^{\rm q} ({\bf k}_{hs}, \omega)$  has a broad peak centered at $\omega =0$.
    In both cases, the spectral function also has a $\delta$-functional peak at $\omega =0$, with overall intensity proportional to the quasiparticle residue~\cite{Berg2023}
   \footnote{
    This $\delta-$function is obtained by either adding $i0$ to $\omega$ (this corresponds to treating $T=0$ as the limit $T \to 0$ and using the fact that at any finite $T$, Im $\Sigma (\kv, 0)$ is finite), or by evaluating $A^{\rm q} ({\bf k}, \omega)$ at ${\bf k}$, for which $\epsilon_{{\bf k} + {\bf Q}}$ is finite, and taking the limit ${\bf k} \to {\bf k}_{hs}$.}.

\begin{figure}
\includegraphics[width=0.75\columnwidth]{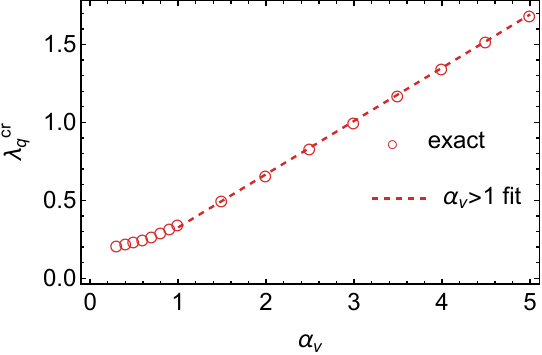}
\caption{Critical $\lambda_{\rm q}$ at different $\alpha_v=v_F/v_s$ and $\theta=\pi/2$, i.e. $\gamma=4 \bar{g}/(\pi v_F^2)$. The spectral function shows pseudogap behavior for $\lambda_{\rm q}<\lambda_{\rm q}^{cr}$. Red circles are $\lambda_{\rm q}^{cr}$,
extracted from the exact formula for the one-loop self-energy, and
 dashed line in the linear fit by
 $\lambda_{\rm q}^{cr}=0.085+\mathsf{c} \alpha_v$ with the same prefactor for $\alpha_v$  that we obtained analytically
 in the $\alpha_v \gg 1$ limit (see the text).
\label{fig:gammacr}}
\end{figure}
 We analyzed the evolution of the spectral function with increasing $\lambda_{\rm q}$ at various
     $\alpha_v = v_F/v_s$
     and found self-consistently critical $\lambda^{cr}_{\rm q}$, at which pseudogap behavior at $T=0$ disappears.
     We show the results in Fig.~\ref{fig:gammacr}.
     For generic $\alpha_v = O(1)$,  $\lambda^{cr}_{\rm q} = O(1)$, i.e., there is no pseudogap in the strong coupling regime. The situation changes when
      $v_s \ll v_F$, i.e.,  $\alpha_v$ is large.  In this limit, we find analytically
      $\lambda^{cr}_{\rm q} = \mathsf{c}\, \alpha_v$, where $\mathsf{c} \approx  (3\sin{\theta}/16)\sqrt{10/3}$.
 Still, at large enough $\xi$, $\lambda_{\rm q}>\lambda^{cr}_{\rm q}$, which implies that near a QCP the one-loop spectral function does not display pseudogap behavior at $T=0$.
  In other words, pseudogap behavior at $T=0$ is {\it not} a precursor to SDW
  \footnote{
  In principle, there may be another possibility to
    suppress the Landau damping even without requiring $v_s \ll v_F$.
      Namely, if one {\it assumes} that the pseudogap exists and evaluate $\gamma$ using the Green's functions with the pseudogap, one find that $\gamma$ indeed gets reduced.  We don't know, however, whether such a state can be ever reached by approaching a QCP from the paramagnetic state.}.

{\it \bf Summary.}~~~
Previous works have found that  in a metal, whose ground state is antiferromagnetically ordered with ${\bf Q} = (\pi,\pi)$
thermal magnetic fluctuations give rise to pseudogap behavior in some temperature range  above $T_N$, when the spectral function of a hot fermion
 contains two peaks, separated by roughly the same energy as in the antiferromagnetically ordered state. This behavior has been obtained theoretically  by departing from free fermions in a paramagnet and evaluating the dressed fermionic Green's function by summing up infinite series of non-crossed and crossed
diagrams for the fermionic Green's function.  The crossed diagrams describe vertex corrections.  We show that keeping vertex corrections is crucial as the combined contribution from non-crossed diagrams
    broadens the spectral function of a hot fermion, but keeps its maximum at zero frequency.
      We argue therefore that to capture
       the physics of a magnetic
      pseudogap,
      one has to go beyond
      self-consistent one-loop theories, like,  e.g., Eliashberg theory for superconductivity.  This result is relevant for the understanding of the observed reduction of superconducting $T_c$ when superconductivity comes out of a pseudogap phase, as within the Eliashberg theory thermal fluctuations do not affect $T_c$. We expect that similar results hold for incommensurate spin fluctuations.

      We also analyzed potential pseudogap behavior at $T=0$, due to quantum fluctuations, assuming no superconductivity. We found that pseudogap may exist at a finite correlation length $\xi$ and may even extend into the strong coupling regime. Still, this  pseudogap behavior   is not the precursor to the ordered state but rather  the consequence of the fact that when spin fluctuations are weakly damped propagating massive bosons, the spectral function of a hot fermion is strongly reduced below the threshold set by the bosonic mass.  We found that sufficiently close to an
      antiferromagnetic QCP, the spectral function of a hot fermion  does not display pseudogap behavior at $T=0$. Combining this with the result of our earlier work~\cite{Ye2023}
      that thermal fluctuations do not give rise to pseudogap behavior when the ground state is not ordered, we conclude that when the ground state is not magnetically ordered, there is no magnetic pseudogap at any $T$       due to long-range magnetic fluctuations.  A potential pseudogap due to short-range fluctuations in a doped Mott insulator has been analyzed in~\cite{Senechal_2004}.

\begin{acknowledgements}
We thank Leon Balents, Erez Berg, Rafael Fernandes, Antoine Georges, Patrick  Lee, Izabella Lovas, Michael Sadovskii,
Subir Sachdev, J\"org Schmalian, Fedor Simkovic, and particularly Andr\'e-Marie Tremblay for helpful discussions and suggestions. M.Y.\ was supported by the Gordon and Betty Moore Foundation through Grant GBMF8690 to UCSB, by a grant from the Simons Foundation (216179, LB), and by the National Science Foundation under Grant No.\ NSF PHY-1748958. AVC  was supported by the U.S. Department of Energy, Office of
Science, Basic Energy Sciences, under Award No.\ DE-SC0014402.
\end{acknowledgements}

\bibliography{PG_thermal}

\clearpage

\appendix 
\onecolumngrid

\section*{Supplemental Material}

\section*{A: Details of self-consistent one-loop theory}
In this section, we show that the pseudogap behavior does not develop in the self-consistent one-loop theory (SCOLT) for any $\lambda_{\rm th}$.

The spectral function in SCOLT is $G^{sc} (\kv_{\rm hs}, \mathsf{\bar w})^{-1} =  v_F \xi^{-1}X$, where
 $\mathsf{\bar w} = {\bar \omega}/v_F \xi^{-1}$, and  $X = X(\mathsf{\bar w})$ is the solution of
  \beq
  X = \mathsf{\bar{w}} - \lambda_{\rm th} \frac{\log\left(X + \sqrt{X^2 +1}\right)}{\sqrt{X^2+1}}
 + \iu \lambda_{\rm th} \frac{\pi}{2\sqrt{X^2+1}}
\label{ff_1_1}
\eeq
(see the main text).
 A way to see whether pseudogap behavior develops is to expand the spectral function $A(\omega) = (1/\pi) \abs{\Im G(\omega)}$ at small $\omega$ as $A(\omega) = A(0) (1 - d\, \omega^2)$ and check the sign of $d$.  Pseudogap develops when $d$ is negative.

 Expanding $X(\mathsf{\bar w})$ at small $\mathsf{\bar w}$, we find
  $X = a\mathsf{\bar w} + ib (1 - c \mathsf{\bar w}^2) +...$, where dots stand for higher-order terms that do not contribute to $d$.
  Solving self-consistently for $a, b$, and $c$, we find after tedious but straightforward algebra  the set of equations
  \bea
 && b \sqrt{1-b^2} = \lambda_{\rm th} \left(\frac{\pi}{2} + i \log{i b + \sqrt{1-b^2}}\right)
 \label{ex_7} \\
 && a = 1 +  \frac{a}{1-b^2} \left (\lambda_{\rm th} - b^2\right) \nonumber \\
 &&c =  \frac{a^2}{2 (1-b^2)} \frac{ 1+2b^2 -3 \lambda_{\rm th}}{1-2b^2 +\lambda_{\rm th}} \nonumber
 \eea
  These equations are valid for $b <1$, which holds for $\lambda_{\rm th} <1$, as we find a'posteriori.

 Evaluating $A(\mathsf{\bar w}) \propto \Im X^{-1} (\mathsf{\bar w})$, we find
 \beq
 d = d_0 \left(1 - \frac{b^2 c}{a^2}\right)
 \eeq
  where $d_0 >0$.  Hence the sign of $d$ is the same as of
  \beq
 1 - \frac{b^2 c}{a^2} = 1 - \frac{b^2( 1+2b^2 -3 \lambda_{\rm th})}{2 (1-b^2) (1-2b^2 +\lambda_{\rm th})}
  \eeq
 where $b$ is determined from (\ref{ex_7}).  Expressing $b = \cos {\psi}$, $\sqrt{1-b^2} = \sin{\psi}$,
 such that
 $\psi =\pi/2$ at $b=0$ and $\psi \to 0$ at $b \to 1$,
  we obtain from (\ref{ex_7}),
  \beq
  2 \lambda_{\rm th} \psi = \sin{2\psi}
 \label{ex_8}
  \eeq
   and
   \beq
   1 - \frac{b^2 c}{a^2} = 1 - \frac{\cos^2 {\phi}}{2\sin{\phi}}~\frac{2 + \cos{2\phi} -3 \lambda_{\rm th}}{\lambda_{\rm th} - \cos{2\psi}}
 \label{ex_9}
   \eeq
 Solving (\ref{ex_8}) for $\psi = \psi(\lambda_{\rm th})$ and substituting into  (\ref{ex_9}), we find that $1 - b^2 c/a^2$
   remains positive for all $\lambda_{\rm th} <1$.  At $\lambda_{\rm th} \to 0$, $1 - b^2 c/a^2 \to 1$, at  $\lambda_{\rm th} \to 1$, $1 - b^2 c/a^2 \to 0.9$.

   For $\lambda_{\rm th} >1$, similar analysis yields
   \beq
   1 - \frac{b^2 c}{a^2} = 1 - \frac{b^2( 1+2b^2 -3 \lambda_{\rm th})}{2 (b^2-1) (2b^2-1-\lambda_{\rm th})}
  \label{ex_10}
   \eeq
 where $b$ is determined by
  \beq
 b \sqrt{b^2-1} = \lambda_{\rm th} \log{b + \sqrt{b^2-1}}
 \label{ex_11}
 \eeq
   Introducing $b = \cosh{\psi}$ and $\sqrt{b^2-1} = \sinh{\psi}$ such that $\psi =0+$ at $b =1+0$ and
   $\psi \approx \log{2b}$ at $b \gg 1$,  we rewrite (\ref{ex_10}) and (\ref{ex_11}) as
    \beq
 \sinh{2\psi} = 2 \lambda_{\rm th} \psi
 \label{ex_11_1}
 \eeq
   \beq
   1 - \frac{b^2 c}{a^2} = 1 - \frac{\coth^2{\psi}}{2}  \frac{ 1+2 \cosh^2{\psi} -3 \lambda_{\rm th}}{2\cosh^2{\psi}-1-\lambda_{\rm th}}
  \label{ex_10_1}
   \eeq
   Solving (\ref{ex_11_1}) for $\psi = \psi (\lambda_{\rm th})$ and substituting into (\ref{ex_10}), we obtain
 that $1 - b^2 c/a^2$ remains positive. Hence, $d \propto (1 - b^2 c/a^2)$ is positive for all $\lambda_{\rm th}$. A positive $d$ implies that the spectral function has a maximum at $\omega =0$, hence pseudogap behavior does not develop.

 \section*{B: Toy models}
 
In the section, we discuss two  toy models for the pseudogap, which both interpolate between SCOLT and  perturbative one-loop theory.

\begin{figure}
\includegraphics[width=0.4\columnwidth]{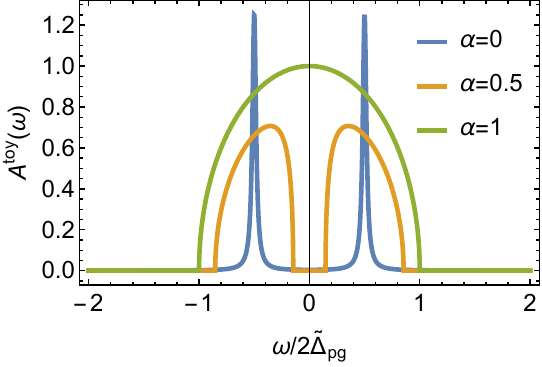}
\caption{Electron spectral function from the toy model of Eq.~\eqref{eq:StoySM}.
\label{fig:AscSM}}
\end{figure}

First, we consider the toy model introduced in the main text (Eq.~(13)). 
 Within this model, the   self-energy at large $\lambda_{\rm th}$ is given by
\beq
\Sigma^{\rm toy1} (\kv_{\rm hs}, \omega) = \tDpg^2 \left(\alpha G(\kv_{\rm hs}, \omega)  + (1-\alpha) G^{(0)}(\kv_{\rm hs}, \omega)\right)
\label{eq:StoySM}
\eeq
where $0 \leq \alpha \leq 1$.
This self-energy interpolates between perturbative one-loop theory at $\alpha =0$ and  SCOLT at $\alpha =1$.  The spectral function $A^{\rm toy} (\kv_{\rm hs}, \omega)$ is readily obtained by solving  the self-consistent equation for the Green's function
 $ G^{-1}(\kv_{\rm hs}, \omega) = \omega - \tDpg^2 \left(\alpha G(\kv_{\rm hs}, \omega)  + (1-\alpha) G^{(0)}(\kv_{\rm hs}, \omega)\right)$ and is shown in Fig.~\ref{fig:AscSM}.
 $A^{\rm toy1} (\kv_{\rm hs}, \omega)$
 is non-zero in a finite range between $|\omega_{min}| = \tDpg (1-\sqrt{\alpha})$ and $|\omega_{max}| = \tDpg (1+\sqrt{\alpha})$. The maxima are at $|\omega|= \tDpg (1-\alpha)^{1/2}$, which remains finite as long as $\alpha <1$. In particular, for $\alpha= 0$, $A^{\rm toy1} (\kv_{\rm hs}, \omega)$ has  $\delta$-functional peaks at $\omega = \pm \tDpg$. For $\alpha =1$ (SCOLT), it is a semi-circle at $|\omega| <2\tDpg$  with the maximum at $\omega =0$, and there is no pseudogap.
We see that the SCOLT is the boundary case for the pseudogap formation.

Another way to interpolate between the one-loop perturbation theory 
and SCOLT
is to consider the self-energy at a hot spot in the form
\beq
\Sigma^{\rm toy2} (\omega) = \frac{\tDpg^2}{\omega - \alpha \Sigma^{\rm toy2} (\omega)}
\eeq
where  $\tDpg = (v_F \xi^{-1}) (\lambda_{\rm th} \log{\lambda_{\rm th}})^{1/2}/\sqrt{2} \approx \left(\frac{3{\bar g} T}
 {2\pi}  \log{\xi}\right)^{1/2}$ is the same as in the main text.   At $\alpha =0$, $\Sigma^{\rm toy2} (\omega)$ is the same as in the one-loop perturbation theory, and the spectral function has two $\delta$-functional peaks at
 $\omega = \pm \tDpg$. At $\alpha =1$, $\Sigma^{\rm toy2} (\omega)$ is the same as in the SCOLT, and the
  the spectral function $A^{\rm toy2}(\omega) \propto \sqrt{4\tDpg^2 - \omega^2}$, with the maximum at $\omega =0$.

The analysis at intermediate $\alpha$ is straightforward, one just has to solve the quadratic equation for
$\Sigma^{\rm toy2} (\omega)$ and substitute the result into the spectral function.  We depart from  $\alpha =1$ and
 gradually decrease $\alpha$.

 For the retarded fermionic Green's function at a hot spot we obtain for $\omega >0$,
 \beq
 G^{\rm toy2}(\omega) = \frac{2\alpha}{\omega (2\alpha -1)+ \sqrt{\omega^2 -4 \tDpg^2 \alpha + i0}}.
 \label{eq:Gtoy2}
 \eeq

\begin{figure}[t]
\subfigure[]{\includegraphics[width=0.25\columnwidth]{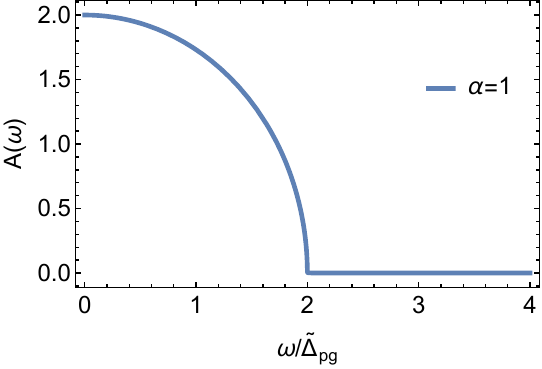}}
\subfigure[]{\includegraphics[width=0.24\columnwidth]{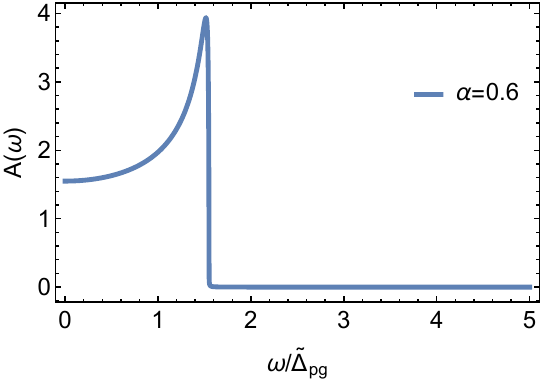}}
\subfigure[]{\includegraphics[width=0.24\columnwidth]{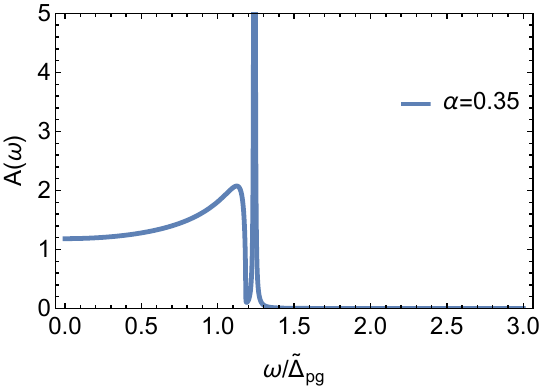}}
\subfigure[]{\includegraphics[width=0.24\columnwidth]{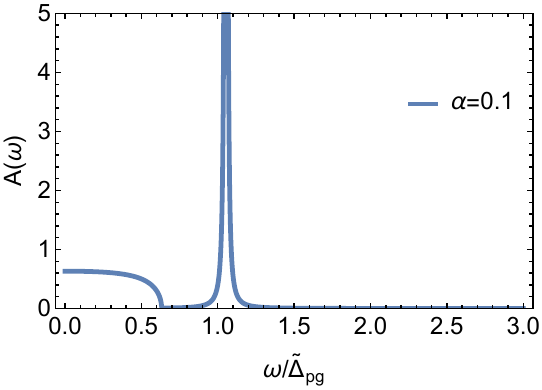}}
\caption{Spectral function at different $\alpha$ for the toy model Eq.~\eqref{eq:Gtoy2}. Pseudogap develops for $\alpha<\alpha_{cr, 1}=0.854$.
\label{fig:Atoy2}}
\end{figure}

A way to check whether pseudogap develops is to expand the spectral function $A^{\rm toy2}(\omega) = (1/\pi) \abs{\Im G(\omega)}$ near $\omega =0$ and check the  sign of the slope. Expanding to order $\omega^2$,  we obtain
 \beq
A^{\rm toy2}(\omega) = \frac{\sqrt{\alpha}}{\pi \tDpg} \left(1 - \frac{\omega^2}{8 \tDpg^2 \alpha} \left(1 -8 \alpha + 8 \alpha^2\right)\right)+O(\omega^4)
\eeq
The maximum of $A^{\rm toy2}(\omega)$ is at $\omega =0$ at $\alpha =1$, and an elementary analysis shows that this holds
 at $\alpha > \alpha_{cr,1} = (1 + 1/\sqrt{2})/2 =0.854$. For these $\alpha$, the  maximum of $A^{\rm toy2}(\omega)$ remains at $\omega =0$, as in SCOLT.  At smaller $\alpha$, the maximum shifts to finite $\omega$, i.e., the spectral function develops a pseudogap behavior. We show this in panels (a) and (b) of Fig.~\ref{fig:Atoy2}.

This behavior holds at $\alpha >1/2$. At smaller $\alpha$, the spectral function develops a $\delta$-functional peak at $\omega = \tDpg /\sqrt{1- \alpha}$ (Fig.~\ref{fig:Atoy2} (c)). The spectral function at these $\alpha$ consists of a $\delta-$function and a continuum at $\omega < 2 \tDpg \sqrt{\alpha}$.  The spectral function in the continuum remains peaked at a finite $\omega$ down to $\alpha_{cr,2} = (1 - 1/\sqrt{2})/2 =0.146$.  At $\alpha < \alpha_{cr,2}$, the continuum part of $A^{\rm toy2}(\omega)$ is peaked at $\omega =0$ (Fig.~\ref{fig:Atoy2} (d)).  Finally, at $\alpha =0$, the continuum disappears, and  $A^{\rm toy2}(\omega)$ only has a $\delta$-functional peak at $\omega = \tDpg$, as in the perturbative one-loop theory.

 \section*{C: Self-energy at $T=0$.}
We use the model of fermions with $t-t'$ dispersion, coupled by Yukawa ${\bar g}$ to dressed
dynamical spin fluctuations with the propagator (in Matsubara frequencies)
$\chi (\ve{q},\Omega_m)=\chi_0/((\Omega_m/v_s)^2+(\ve{q}-\ve{Q})^2+\xi^{-2} + \gamma |\Omega_m|)$, where $v_s$ is spin velocity, ${\bf Q} = (\pi,\pi)$, and the Landau damping term $\gamma |\Omega_m|$ comes from inserting particle-hole bubbles into the spin propagator using the same Yukawa spin-fermion coupling.
 The prefactor $\gamma$ then scales with ${\bar g}$ and is given by
 ~\cite{Abanov2003}
 \beq
 \gamma = \frac{4}{\pi \sin{\theta}} \frac{{\bar g}}{v^2_F},
 \eeq
 where $\theta$ is the angle between the directions of Fermi velocities at ${\bf k}_{hs}$ and ${\bf k}_{hs} + {\bf Q}$.

 The one-loop self-energy is the convolution of the bare fermionic Green's function $G_0 ({\bf k}_{hs} + {\bf q}, \omega_m + \Omega_m)$ and $\chi ({\bf q},\Omega_m)$.  Expanding the dispersion to linear order in ${\bf q} - {\bf Q}$ and performing angular integration, we obtain
\begin{align}
\Sigma(\kv,i \mathsf{w}_m) \approx -i v_F \xi^{-1} \frac{\lambda_{\rm q}}{2\pi} \int \diff \mathsf{W}_m \sgn (\mathsf{w}_m + \mathsf{W}_m) \int\frac{\diff z}{\sqrt{z+ (\mathsf{w}_m+ \mathsf{W}_m)^2} (z+1+ \alpha_v^2 \mathsf{W}^2_m + {\bar \gamma} |\mathsf{W}_m|)}
\label{eq:Eliashberg0}
\end{align}
where
\beq
\lambda_{\rm q} = \frac{3}{4\pi} \frac{ {\bar g}}{v_F \xi^{-1}}
\eeq
and we introduced  $\mathsf{w}_m = \omega_m/(v_F \xi^{-1})$,  $\mathsf{W}_m = \Omega_m/(v_F \xi^{-2})$, ${\bar \gamma} = \gamma v_F \xi = \frac{16}{3 \sin\theta}\lambda_{\rm q}$, and $\alpha_v = v_F/v_s$.  At small $\lambda_{\rm q}$, Landau damping is weak.
 Neglecting it, evaluating the frequency integral, and converting from Matsubara to real axis, we find
  that  Im $\Sigma(\kv, \mathsf{w})$ vanishes at $|\mathsf{w}| <1/\alpha_v$. At larger $|\mathsf{w}|$,
  Im $\Sigma(\kv, \mathsf{w})$ is non-zero.  Substituting $\Sigma(\kv, \mathsf{w})$  into the expression for the spectral function $A^{\rm q}
   ({\bf k}_{hs}, \omega)$, we find that  it has a maximum at $|\mathsf{w}| \geq 1/\alpha_v$, i.e., at
     $|\omega| \geq v_s \xi^{-1}$.
   In the opposite limit of large $\lambda_{\rm q}$, the Landau damping term is the strongest one in the spin propagator,  and performing the same calculation we find that $A^{\rm q} ({\bf k}_{hs}, \omega)$  has a broad peak centered at $\omega =0$.

  At large $\alpha$, typical $z$ in (\ref{eq:Eliashberg0}) are parametrically larger than $(\mathsf{w}_m+ \mathsf{W}_m)^2$ (typical $\mathsf{W}_m$ is of order $\mathsf{w}_m$).  Neglecting $(\mathsf{w}_m+ \mathsf{W}_m)^2$
   and integrating over $z$, we obtain
   \beq
\Sigma(\kv,i \mathsf{w}_m) \approx -i v_F \xi^{-1}\lambda_{\rm q} \int_0^{\mathsf{w}_m}
 \frac{d \mathsf{W}_m}{\sqrt{1+ \alpha_v^2 \mathsf{W}^2_m + {\bar \gamma} |\mathsf{W}_m|}}
\label{eq:Eliashberg1}
\eeq
Evaluating the remaining frequency integral, converting the self-energy onto the real axis, and evaluating the spectral function, we
 find analytically that pseudogap exists as long as $\lambda_{\rm q} < \lambda_{\rm q}^{cr}$, where the latter is the solution of
 \beq
 \lambda_{\rm q}^{cr} = \frac{3\sin{\theta}}{16} \frac{2}{\sqrt{3}} \sqrt{\frac{9-4 \lambda^*}{5-4 \lambda^* + (\lambda^*)^2}} \alpha_v.
 \label{eq:Eliashberg2}
 \eeq
 where $\lambda^* = \lambda_{\rm q}^{cr}/(1+\lambda_{\rm q}^{cr})$.  Solving this equation, we find, at large $\alpha_v$,
 $\lambda^{cr}_{\rm q} = \mathsf{c} \alpha_v$, where $\mathsf{c} \approx  (3\sin{\theta}/16)\sqrt{10/3}$. We presented this result in the main text.

\begin{figure}
\includegraphics[width=0.45\columnwidth]{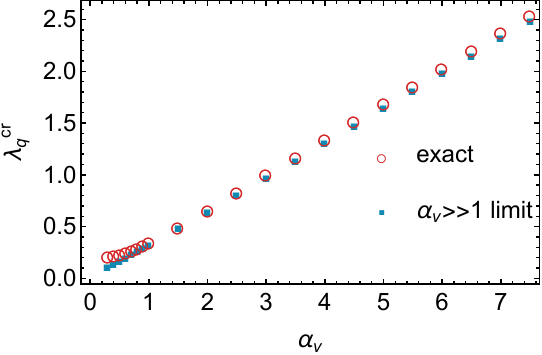}
\caption{Critical $\lambda_{\rm q}$ at different $\alpha_v=v_F/v_s$ for $\theta=\pi/2$. The spectral function shows pseudogap behavior for $\lambda_{\rm q}<\lambda_{\rm q}^{cr}$. Red circles are $\lambda_{\rm q}^{cr}$,
extracted from the exact formula for the one-loop self-energy, Eq.~\eqref{eq:Eliashberg0}, and
blue dots are obtained by using the approximate formula, Eq.~\eqref{eq:Eliashberg1}, i.e., by solving Eq.~\eqref{eq:Eliashberg2}.
\label{fig:gammacr_1}}
\end{figure}

In Fig. \ref{fig:gammacr_1} we compare $\lambda_{\rm q}^{cr}$, obtained by using the full expression for the self-energy, Eq. (\ref{eq:Eliashberg0}), and by using the approximate Eq. (\ref{eq:Eliashberg1}), which we justified at large $\alpha_v$. The same expression as (\ref{eq:Eliashberg1}) is obtained by factorizing the momentum integration along and transverse to the Fermi surface. This factorization is in turn justified if bosons are slow modes compared to fermions.
 We see from Fig. \ref{fig:gammacr_1} that the values of $\lambda_{\rm q}^{cr}$, obtained using the exact and the approximate forms of the self-energy, are essentially identical for all $\alpha_v \geq 1$.
 
\end{document}